\documentclass[12pt,preprint]{aastex}
\usepackage{emulateapj5}
\usepackage{onecolfloat}
\usepackage{epsf}

\newenvironment{inlinefigure}{
\def\@captype{figure}
\noindent\begin{minipage}{0.999\linewidth}\begin{center}}
{\end{center}\end{minipage}\smallskip}

\newcommand{\sirtf}{\textit{SIRTF}}
\newcommand{\iras}{\textit{IRAS}}
\newcommand{\iso}{\textit{ISO}}

\newcommand{\AAA}{\hbox{\AA}}

\newcommand{\lsim}{\lesssim}
\newcommand{\gsim}{\gtrsim}

\newcommand{\etal}{et al.}
\newcommand{\eg}{e.g.}
\newcommand{\ie}{i.e.}

\shorttitle{On the IR Luminosity of Distant Galaxies}
\shortauthors{Papovich \& Bell}
\slugcomment{To Appear in the Astrophysical Journal Letters, 2002
November 1}

\begin{document}

\def\head{

\title{On Measuring the Infrared Luminosity of Distant Galaxies\\ 
 with the \textsl{Space Infrared Telescope Facility}}

\author{Casey Papovich and Eric F.\ Bell\altaffilmark{1}}
\affil{Steward Observatory, The University of Arizona, 933
	N. Cherry Avenue, Tucson AZ 85721, USA;}
\email{papovich@as.arizona.edu, ebell@as.arizona.edu}



\begin{abstract}

The {\it Space Infrared Telescope Facility} (\sirtf) will
revolutionize the study of dust--obscured star formation in distant
galaxies.  Although deep images from the Multiband Imaging Photometer
for \sirtf\ (MIPS) will provide coverage at 24, 70, and 160\micron,
the bulk of MIPS--detected objects may only have accurate photometry
in the shorter wave\-length bands due to the confusion noise.
Therefore, we have explored the potential for constraining the total
infrared (IR) fluxes of distant galaxies with solely the 24\micron\
flux density, and for the combination of 24\micron\ and 70\micron\
data.  We also discuss the inherent systematic uncertainties in making
these transitions.  Under the assumption that distant star-forming
galaxies have IR spectral energy distributions (SEDs) that are
represented somewhere in the local Universe, the 24\micron\ data
(plus optical and X-ray data to allow redshift estimation and AGN
rejection) constrains the total IR luminosity to within a factor of
2.5 for galaxies with $0.4 \lsim z \lsim 1.6$.  Incorporating the
70\micron\ data substantially improves this constraint by a factor
$\lsim 6$.  Lastly, we argue that if the shape of the IR SED is known
(or well constrained; \eg, because of high IR luminosity, or low
ultraviolet/IR flux ratio), then the IR luminosity can be estimated
with more certainty.
\end{abstract}
 
\keywords{
cosmology: observations ---
galaxies: high-redshift ---
infrared: galaxies
}
}

\twocolumn[\head]
\altaffiltext{1}{Present Address: Max Planck Institut f\"ur
Astronomie, K\"onigstuhl 17, D-69117 Heidelberg, Germany; bell@mpia.de}


\section{Introduction}

The evolution of the global, volume-averaged star formation rate (SFR)
is currently a topic of intense interest \citep[see,
e.g.,][]{mad96,ste99,yan99,haarsma00}.  Observationally, a galaxy's
SFR must be inferred from the luminosity density at wave\-lengths
dominated by young stars, such as the ultraviolet (UV), nebular
emission lines, mid--to--far infrared (IR), or radio \citep[see,
\eg,][]{ken98a,con92}, and by making assumptions about the form of the
IMF.  Because of dust, rest--frame UV/optical indicators often
underestimate the intrinsic SFR, especially for more luminous galaxies
\citep[see, e.g.,][]{cal94,cal01,bel02b}.  Based on the UV/optical
spectral energy distributions (SEDs) and far--IR emission of distant
galaxies and QSOs, dust appears to be prevalent in  high--redshift
objects \citep[$z\sim 2-5$, \eg,][]{pet98,ade00,car00,pap01,cha02}.
Dust reprocesses the energy absorbed from the UV/optical into the
mid/far--IR, and thus galaxy IR luminosities allow one to `balance the
energy budget' and paint a much more complete picture of star
formation throughout cosmic history \citep[see,
e.g.,][]{san96,bla99,flo99}.

The {\it Space Infrared Telescope Facility} (\sirtf), to be launched
in 2003,  offers to revolutionize our view at IR
wave\-lengths with unprecedented sensitivity and resolution.  Several
cosmological surveys  are being planned to obtain 24, 70, and
160\micron\ data with the Multiband Imaging Photometer for \sirtf\
\citep[MIPS; see, e.g.,][]{lon01,dic01,rie01,rie01a}.  The combination
of these three bands will span the peak of the IR SEDs and provide a
fairly robust tracer of the total IR emission \citep[\eg,][]{dal02}.
However, MIPS observations will rapidly become confusion limited due
to the increasing source density at faint fluxes (\eg, Xu \etal\ 2001;
Dole, Lagache, \& Puget 2002), especially for the longer wave\-length
data, which have lower sensitivity and resolution (MIPS resolution is
roughly proportional to the bandpass central wave\-length).  Thus,
full coverage of galaxies' IR SEDs with accurate photometry may only
be possible for relatively nearby or bright objects.

In this {\it Letter} we consider the uncertainties inherent in
translating MIPS photometry into total IR fluxes in the case that
object photometry is only available from the shorter wave\-length
bands.  The relation between mid-IR and far--IR is complex.  Thus we
explore the connection between the 24\micron\ data and the total IR
emission from both the observational (\S2) and modeling (\S3)
perspective, with the assumption that local galaxy SEDs are
representative of high--redshift analogs \citep[which has not been
robustly demonstrated, although see][]{ade00,elb02}.  In \S4, we
investigate improving these constraints using 70\micron\ data, the
galaxy luminosity, and/or UV flux.  We summarize our results in
\S\ref{sec:conc}.  Because extensive X-ray/optical/near-IR coverage
will be available for the majority of deep MIPS survey regions (see
references above), we assume here that galaxies with dominant AGN
contributions can be rejected using the X-ray data, and that galaxies'
redshifts (spectroscopic or photometric) will be known.

\section{Observational constraints between the mid--IR and total IR
emission: the case at $z\sim 1$}\label{sec:emp} 

To explore the correlation between mid--IR fluxes and the total IR
flux\footnote{We define the total IR flux as the integrated emission
from $8-1000$\micron, \ie, $F_\mathrm{IR} \equiv F(8-1000\micron)$,
estimated using direct integration from $8-100$\micron\ and
extrapolated to 1000{\micron} using a $\lambda^{-1}$ emissivity.
These fluxes are typically a factor of 2 larger than those of
\citet{hel88} and 30\% higher than those of  \citet{san96}, and are
consistent to within 10\% with values extrapolated using
8--170{\micron} data from \iras\ and \citet{tuf02}.} for local
galaxies, we have taken data from the {\it Infrared Astronomical
Satellite} (\iras) at 12\micron.  
\begin{inlinefigure}
\vspace{-1.0cm}
\hspace{-0.5cm}
\begin{center}
\resizebox{ 1.0\textwidth}{!}{\includegraphics{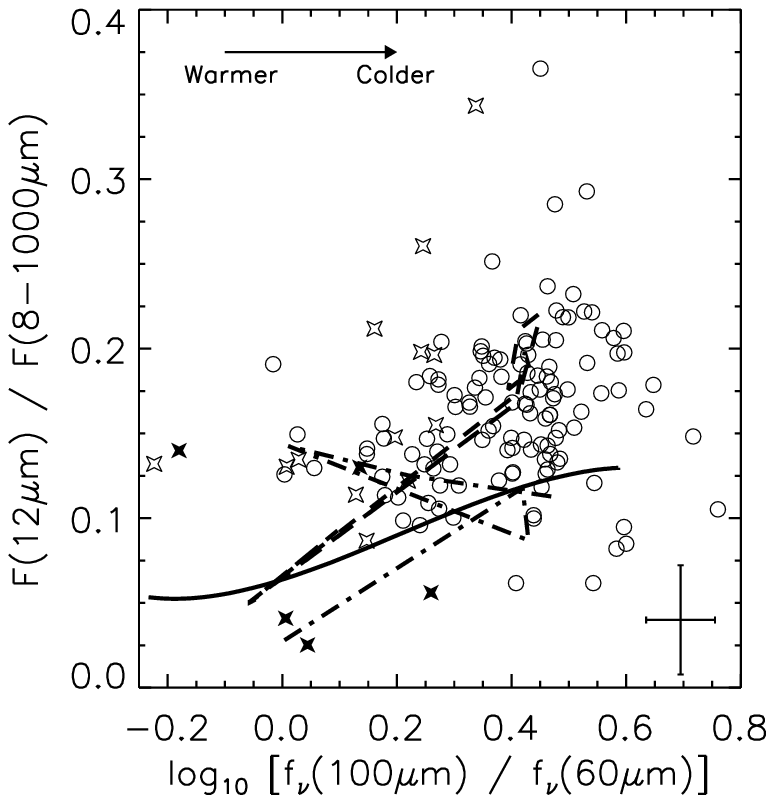}}
\end{center}
\vspace{-0.5cm}  \figcaption{The ratio of the \iras\ 12\micron\ to
total IR flux (from $8-1000$\micron) as a function \iras\ 100\micron\
to 60\micron\ flux--density ratio; see \citet{bel02b} for details.
Symbols denote various galaxy types: starbursting galaxies
(\textsl{open stars}); ultra--luminous IR galaxies (\textsl{filled
stars}); normal galaxies  (\textsl{open circles}). The error bar
indicates typical uncertainties.  Results for model galaxies are shown
as lines (with line definitions as in figure~\ref{fig:24umvz}).
\label{fig:iras12um}\vspace{0.2cm}}
\end{inlinefigure}
Note that the MIPS 24\micron\ band
in the observed frame for $z=1$ nearly matches the rest--frame \iras\
12\micron\ band, with a $K$--correction that varies weakly with
galaxy SED:
$L_\nu(\mathrm{MIPS}\,24\micron,\,z=1) \approx 1.3\, L_\nu(\mathrm{\it
IRAS}\, 12\micron,\, z=0)$.

Figure~\ref{fig:iras12um} shows the \iras\ 12\micron\ to total IR flux
ratio as a function of \iras\ 100\micron\ to 60\micron\ color for 156
galaxies spanning all star--forming types \citep{bel02b}.\footnote{We
denote the average flux density measured from a  bandpass with central
wave\-length $\lambda$ as $f_\nu(\lambda)$, and we define the flux
from the bandpass as $F(\lambda) \equiv \nu\,f_\nu(\lambda)$, where
$\nu$ is the frequency corresponding to the wave\-length $\lambda$.}
The figure illustrates that a nearly constant proportion of the total
IR flux emerges in the mid-IR for local galaxies.  The mean ratio is
$F(12\micron)/F_\mathrm{IR} = 0.16$, with a  $1\sigma$ scatter between
0.10 and 0.22.  This is consistent with \iras\ and \iso\ results,
which show a relatively constant conversion between rest-frame mid-IR
($7\micron - 15\micron$) and total IR luminosities
\citep[\eg,][]{spi95,cha01,rou01}.  Clearly, in local galaxies the
dust responsible for the unidentified IR bands \citep[UIBs; most
plausibly associated with polycyclic aromatic hydrocarbons (PAHs),
e.g.,][]{pug89,gen00} that dominates the flux in the mid--IR
correlates fairly well with the dust that produces the bulk of the
total IR flux emission.

Importantly, figure~\ref{fig:iras12um} shows that the ratio of
12\micron\ to total IR flux does not sensitively depend on the temperature
of the dust that dominates at far--IR wave\-lengths (as measured by
$f_\nu(100\micron)/f_\nu(60\micron)$): the systematic variation with
dust temperature seems to be smaller than the scatter.  Overlaid in
the figure are the results of models from the literature with
different treatments of the mid/far-IR SEDs and that span a range of
dust temperatures and IR luminosities (described in \S3).   These
models broadly reproduce the mean and scatter observed in the plot.
These \iras\ data suggest that if the relationship between the mid--IR
and total IR emission at high--redshifts is similar to that observed
locally, then the MIPS 24\micron\ data will provide an estimate to the
total IR emission for $z\sim 1$ galaxies.  Clearly, the scatter is
significant (a factor of $\approx 2.5$), and is largely systematic in
the sense that it can only be improved upon by better constraining the
shape of the galaxy IR SED.

\section{Modeling the mid--IR to total IR emission: extending the
redshift range} \label{sec:model} 

Here, we turn to models to test the result from \S2 for a wider range
of redshifts.  In order to gauge the systematic model uncertainties,
we explore three independently constructed sets of models.  {\it i})
We use the {\sc stardust} templates for 17 local galaxies with a range
of IR luminosities, $5\times 10^9 \lsim L_\mathrm{IR} / L_\odot \lsim
4\times 10^{12}$, from Devriendt, Guiderdoni, \& Sadat (1999).  Note
that the luminosities of the seven Virgo Cluster galaxies in this
model were underestimated by a factor of between 10 and 300
(J.~Devriendt, 2002, private communication); this does not affect our
results as we use only colors (flux ratios).  {\it ii}) We have also
used the fits of the {\sc grasil} model of \citet{sil98} to six
galaxies (Arp~220, M51, M82, M100, NGC~6090, NGC~6946).  {\it iii})
Lastly, we use the empirical models based on local galaxy \iso\ data
from \citet{dal01} with a range of far--IR flux ratios (a surrogate
for dust temperature), $-0.4 \lsim \log F(60\micron)/F(100\micron)
\lsim 0.5$.  These models apply explicitly to local galaxy SEDs, and
thus the caveat persists that they may not describe high--$z$ galaxies.

To derive the observed--frame mid-IR flux as a function of redshift,
we averaged the model galaxy SEDs with the MIPS 24\micron\ bandpass
(including effects from the filter transmission and detector response)
over a range of redshifts.  In figure~\ref{fig:24umvz}, we show the
ratio of the observed-frame MIPS 24\micron\ to rest-frame total IR
flux versus redshift.   As expected from the initial analysis
of 12{\micron} \iras\ data, the MIPS 24\micron\ emission is a fair
tracer of the total (rest-frame) IR emission, especially for the redshift
range $0.4 \lsim z \lsim 1.6$, for which the scatter is a factor of
$2-3$.

It is worth discussing this in somewhat more detail.   At $z \sim 0$,
the observed 24{\micron} band traces emission from UIBs, and very
small grains that have been transiently raised to high temperatures by
absorption of single UV photons \citep{des90}.  Model SEDs show great
variation in the prominence of this component, depending largely on
the relative intensity of the UV radiation field in a given galaxy.
As one moves toward $z \sim 1-2.5$, the MIPS 24\micron\ data probes
lower rest--frame wave\-lengths where the UIBs increasingly contribute
to the flux especially in discrete features from $\sim 7-13$\micron\
\citep[and references therein]{gen00}.  Empirically, these UIBs are
rather more prominent for `cooler' galaxies with less intense UV
radiation fields (causing the tracks for warmer and cooler galaxies to
cross at $z \sim 0.5$).  The `dip' in 24\micron-to-total IR ratio at
$z \sim 1.4$ is  attributed to gaps between the UIBs.   \citet{sil98},
\citet{dev99}, and \citet{dal01} model the UIBs somewhat differently.
The largest difference between the models at 
\begin{inlinefigure}
\begin{center}
\resizebox{ 0.98\textwidth}{!}{\includegraphics{f2.ceps}}
\end{center}
\figcaption{\label{fig:24umvz} The ratio of the observed MIPS 24~\micron\
emission to total (rest-frame) IR emission  as a function of redshift.
Curves display the models described in the text (labeled in the figure
inset).  The arrow depicts galaxy SEDs from ``Warmer'' to ``Colder''
dust temperatures.\vspace{0.1cm}}
\end{inlinefigure}
$z \gsim 1$ is the
contribution of the stellar continuum to the mid--IR emission.  The
galaxy templates match well for warm galaxies, and diverge for cooler
galaxies, as Dale \etal\ do not have a large stellar contribution in
their cold galaxies, whereas Devriendt \etal\ include seven Virgo
Cluster galaxy templates that are dominated by stellar emission at
$\lambda \la 15$\micron\ \citep{bos98}, and Silva \etal\ study three
quiescently star--forming spirals (M100, M51, NGC~6946) that contain
some stellar contribution to the mid--IR emission.  Certainly, the
contribution of the old stellar population to dust heating is
correlated with dust temperature \citep[see, e.g,][and references
therein]{bel02b}; however,  the scatter is large, leaving the issue of
stellar contamination  of $\lambda \lsim 15$\micron\ fluxes largely
open.  Moreover, how these processes contribute to the IR SEDs of
high--$z$ galaxies is almost entirely unknown.

Given this (and with the inherent assumption that distant star-forming
galaxies have IR SEDs that are represented somewhere in the local
Universe), one expects that at $0.4 \lsim z \lsim 1.6$ it should be
possible to constrain the total IR flux to within a factor of $2-3$
using 24\micron\ data alone: $F(24\micron)$ represents $5-25$\% of the
total IR flux.  Outside this redshift range, the conversion is rather
less certain.   A factor of $\approx 2-3$ in error in total IR flux
may be quite adequate, but can potentially frustrate many scientific
analyses.  Because this uncertainty is largely systematic, one must
include additional knowledge of the shape of the galaxy SED in order
to improve the constraint on the total IR flux.

\section{Improving the total IR emission constraints} \label{sec:dis} 

Based on the models illustrated in figure~\ref{fig:24umvz}, the
scatter between the 24\micron\ flux and the total IR flux emission
stems primarily from a lack of knowledge on the shape of the galaxy IR
SED (and/or dust temperature).  MIPS 
\begin{inlinefigure}
\vspace{-0.2cm}
\begin{center}
\resizebox{ 0.98\textwidth}{!}{\includegraphics{f3.ceps}}
\end{center}
\vspace{-0.2cm} 
\figcaption{The ratio of the observed \sirtf/MIPS 24~\micron\  emission
to the total (rest--frame) IR emission  as a function of MIPS
24\micron/70\micron\ flux ratio.  Curves with symbols illustrate the
various predictions from the models described in the text averaged
over the redshift interval (as listed in the figure inset), with line
styles as indicated in figure~\ref{fig:24umvz}.  The thick solid and
dotted lines show a second-order polynomial fit and its standard error
to the average over all model predictions and redshifts (see text).
\vspace{0.1cm}\label{fig:24umv2470}}
\end{inlinefigure}
70\micron\ flux measurements will
be available for a significant fraction of MIPS 24\micron\ sources
(although the exact fraction and redshift distribution are strongly
model dependent), which should enhance the constraints on the total IR
emission.   
Here we explore the constraints on the total galaxy IR emission in the
case that only MIPS 24\micron\ and 70\micron\ data are available.  In
figure~\ref{fig:24umv2470} we show the ratio of the observed MIPS
24\micron\ to total IR emission versus the ratio of the observed MIPS
24 and 70\micron\ fluxes in different redshift intervals.  There is
generally very little scatter in the relationship for these redshifts
and colors: the ratio of the 24--to--70\micron\ fluxes seems to
correlate with dust temperature over a wide range of redshifts.  To
crudely parameterize the relationship between the IR fluxes in these
bandpasses, we have fit a second-order polynomial to the average over
all models for all redshifts shown in figure~\ref{fig:24umv2470}.  We
find a best fit to the equation, $F(24\micron)/F(8-1000\micron) = A +
Bx + Cx^2$, where $x \equiv \log F(24\micron)/F(70\micron)$, with
polynomial coefficients, $A = 0.160$, $B=0.135$, and $C = 0.033$, and
a standard error of $\delta[F(24\micron)/F(8-1000\micron)] = 0.036$.
This fit and error are overlaid in figure~\ref{fig:24umv2470}.  Our
parameterization provides a means for quantifying the scatter in
estimating the total IR flux emission using the MIPS 24 and 70\micron\
data (although the physical relation between these fluxes and the
total IR emission is of course more complicated than this simple
formalism), and it spans the systematic uncertainties inherent in the
model predictions and is largely independent of redshift between $0.3
\le z \le 3.3$.  Therefore with both MIPS 24 and 70\micron\ detections
(and with the assumption that  local galaxy IR SEDs apply to
high--redshift analogs) this result suggests that the uncertainty on
the total IR flux is improved by up to a factor $\sim 6$ (relative to
that achieved using 24\micron\ data only).

Lastly, if distant luminous IR galaxies (LIRGs; $L_\mathrm{IR} >
10^{11} L_\odot$) have comparable SEDs to those found locally, then it
should be possible to better constrain the IR luminosity for this
class of galaxy.  Locally, LIRGs have ``Warmer'' dust temperatures
\citep{san96}, thus allowing for a more robust conversion between
observed 24um and total IR luminosity (adopting the warmer tracks,
with less stellar contribution, in figure~\ref{fig:24umvz}).  It may
be possible to classify these galaxies as LIRGs on the basis of their
high IR luminosities (indeed, although model dependent, it may be the
case that only LIRGs will be detectable in confusion--limited surveys
for $z\gsim 1$; \eg, Dole \etal\ 2002).  Alternatively, many local
LIRGs have low UV-to-IR ratios $F(1550\AAA)/F_\mathrm{IR} < 0.02$,
which can be derived from observed-frame $U$-band to 24\micron\ ratios
for $z\sim 1$ galaxies, albeit in a model-dependent way.  These
conversions obviously depend sensitively on high redshift LIRGs acting
in similar ways to local LIRGs, which of course must be tested using
spectral observations along with 70 and 160\micron\ photometry for the
brightest galaxies.


\section{Conclusions} \label{sec:conc}

\sirtf\ will greatly improve our understanding of the IR emission of
(and hence the star--formation processes within) distant galaxies.
Because deep surveys could uncover many star--forming galaxies in the
shorter--wave\-length MIPS data with no longer--wave\-length
counterparts, we have explored the efficacy of constraining the total
IR galaxy emission using the data at 24\micron\ only and the
combination of 24 and 70\micron\ (and under the assumption that
ancillary optical and X-ray data are available to allow redshift
estimation and AGN rejection).  Assuming that distant star-forming
galaxies have IR SEDs that are represented somewhere in the local
Universe, the 24\micron\ data should  constrain the integrated IR flux
to within a factor of 2.5: $F(24\micron) / F_\mathrm{IR} \simeq
5-25$\% for galaxies with $0.4 \lsim z \lsim 1.6$.  Including MIPS 
70\micron\ data, the IR luminosity can be estimated with considerably
more certainty (a factor $\lsim 6$ improvement over the uncertainties
for using 24\micron\ data only).  Lastly, if one can assume that the
shape of the galaxy SED is similar to local luminous IR galaxies ($L
\gsim 10^{11}\, L_\odot$; \eg, because of high IR luminosity or low
UV/IR flux ratio), then the constraints should improve.

Throughout this study we have made the explicit assumption that the
relationships between the mid-- and total--IR emission and the shape
of the IR SEDs observed locally apply to properties of high--redshift
galaxies.  As a final caveat, we emphasize that for many reasons this
may not be the case.  Indeed, the population of IR--emitting galaxies
must undergo strong evolution in order to match the cosmic IR
background \citep{hau01,cha01,xu01,elb02,dol02}, and it is unclear how
such evolution manifests itself.
The mid--IR UIBs (\eg, PAHs) and total IR emission relation at high
redshift may be different due to, \eg, a changing dust composition, or
a significantly different dust heating from older stellar populations
than that observed locally.
Moreover, we have assumed that the shape of the IR SED as a function
of IR luminosity for local galaxies applies to high--redshift analogs,
which is poorly known \citep[although not inconsistent with
observations; see, \eg,][]{ade00}.  Only with large samples of
galaxies detected in all three MIPS bands, and with spectroscopic
measurements from the \sirtf\ Infrared Spectrograph (IRS), will these
assumptions be testable.  Once such constraints are established, the
data should provide a prescription for relating the MIPS data to total
IR fluxes.  Given an estimate  on the total IR luminosity, it is then
possible to estimate SFR \citep[modulo the usual sources of
uncertainty; see, e.g.,][]{ken98a,bel02b}.

\acknowledgements

We wish to thank our colleagues at the Steward Observatory for
stimulating conversations, in particular Herv\'e Dole, Rob Ken\-ni\-cutt,
John Mous\-ta\-kas, Mar\-cia Rie\-ke, George Rie\-ke, and J.~D. Smith.  We are
also grateful to D.~Dale, J.~Devriendt, and L.~Silva for providing
their galaxy SEDs, and to the anonymous referee whose comments
improved this work.  C.~P.\ wishes to thank G.~Rieke for kindly
providing support through the \sirtf\ project via JPL contract no.\
960785.  E.~F.~B.\ was supported by NASA grant NAG5-8426 and NSF grant
AST-9900789.


\end{document}